\documentclass[reprint,%
 amssymb, amsmath,%
 groupedaddress,%
]{revtex4-1}

\usepackage{docs,amsfonts,amssymb,amsmath,graphicx,mathrsfs,bm}
\usepackage{subfigure,float,txfonts}
\usepackage{booktabs}
\begin{document}

\title{Transmission comb of a distributed Bragg reflector induced by two surface dielectric gratings}
\author{Xiaobo Zhao}
\author{Bingsuo Zou}
\author{Yongyou Zhang}
\email[Author to whom correspondence should be addressed. Electronic mail: ]{yyzhang@bit.edu.cn}
\affiliation{$^1$Beijing Key Lab of Nanophotonics $\&$ Ultrafine Optoelectronic Systems and School of Physics, Beijing Institute of Technology, Beijing 100081, China}%

\begin{abstract}
With transfer matrix theory, we study the transmission of a distributed Bragg reflector (DBR) with two dielectric gratings on top and on the bottom. Owing to the diffraction of the two gratings, the transmission shows a comb-like spectrum which red shifts with increasing the grating period during the forbidden band of the DBR. The number density of the comb peaks increases with increasing the number of the DBR cells, while the ratio of the average full width at half maximum (FWHM) of the transmission peaks in the transmission comb to the corresponding average free spectral range, being about 0.04 and 0.02 for the TE and TM incident waves, is almost invariant. The average FWHM of the TM waves is about half of the TE waves, and both they could be narrower than 0.1 nm. In addition, the transmission comb peaks of the TE and TM waves can be fully separated during certain waveband. We further prove that the transmission comb is robust against the randomness of the heights of the DBR layers, even when a 15\% randomness is added to their heights. Therefore, the proposed structure is a candidate for a multichannel narrow-band filter or a multichannel polarizer.
\end{abstract}

\maketitle

\noindent Distributed Bragg reflector (DBR) is an important element widely used in optics \cite{Tawara2003Low-APL, Sheinfux2014Subwavelength-PRL, OBrien2014Comb-OL, Zhang2014Efficient-OE,Zhang2009Ultra-short-OE, Kulishov2013Distributed-OE, Dong2012Tailoring-OL}, photonics \cite{Chen2012High-APL, Benson2011Assembly-Nature, Armani2003Ultra-Nature}, solar cells \cite{Sheng2012Integrated-APL}, and so on. Though it is a traditional and well-developed optical structure, the DBR still attracts plenty of attentions due to its high tunability and extensibility which also could be enhanced by introducing additional structures, for example, defects and gratings. These benefits lead to many applications of the DBR, such as optical switches \cite{Anton2012Dynamics-APL, Steger2012Single-APL}, lasers \cite{Gessler2014Low-APL, Chen2006Nanowire-APL},  couplers \cite{Kulishov2013Distributed-OE}, and narrow band filters \cite{Levy-Yurista-APL-2000, OBrien2014Comb-OL, Dong2012Tailoring-OL}. For narrow-band filters, one could turn to the semiconductor microcavity which consists of two distributed Bragg reflectors (DBRs) and one cavity layer \cite{Deng-RMP-2010, Lai-Nature-2007}. When the cavity layer is high enough, the transmission of the semiconductor microcavity is comb-like with the peak separation being roughly inversely proportional to the distance between the two DBRs \cite{OBrien2014Comb-OL}. In this case, the semiconductor microcavity can serve as a multichannel optical filter. The transmission peaks locate in the forbidden band of the DBR, thus the small peak separation in the comb-like transmission is required. As a result, the cavity layer thickness being vertical to the DBR should be much larger than the photon wavelength when one wants to achieve the small peak separation. 

In order to avoid this condition, we consider the optical structure consisted of one DBR and two same dielectric gratings on top and on the bottom, respectively, as shown in Fig.~1(a). With this structure, we plan to achieve the multichannel optical filter, simultaneously with small size. The grating geometry has been applied to many kinds of optical structures to enhance or change the device properties \cite{Pruessner2007Integrated-OL, Sheng2012Integrated-APL, Dong2012Tailoring-OL, Sharon-APL-1996, Levy-Yurista-APL-2000, Zhang-NL-2008}. In the present work, two gratings are introduced to adjust the transmission spectra during the forbidden band of the DBR. The DBR whose wavelength at the gap center is designed to be $\lambda_c=900 $ nm, consists of GaAs and AlAs layers. The heights of GaAs and AlAs are $h_a=\lambda_c/4n_a$ and $h_b=\lambda_c/4n_b$ with $n_a=3.58$ and $n_b=3.0$ being their refractive indexes, respectively. In this work, we assume the DBR has $N$ layers of GaAs and $N+1$ layers of AlAs where $N$ represents the cell number of the DBR, and thus the total layer number is $2N+3$ for the structure in Fig.~1(a). Besides, we assume the GaAs gratings on top and on the bottom have the same height $h_g$, width $L_g$, period $L$, and thus the duty cycle $f_g=L_g/L$. All parameters are marked in Fig. 1(a) for clarity. Without loss of generality, the medium above and below DBR is set to be air whose refractive index is $n_0=1$.

\begin{figure}[htb]
\centerline{\includegraphics[width=7.5cm]{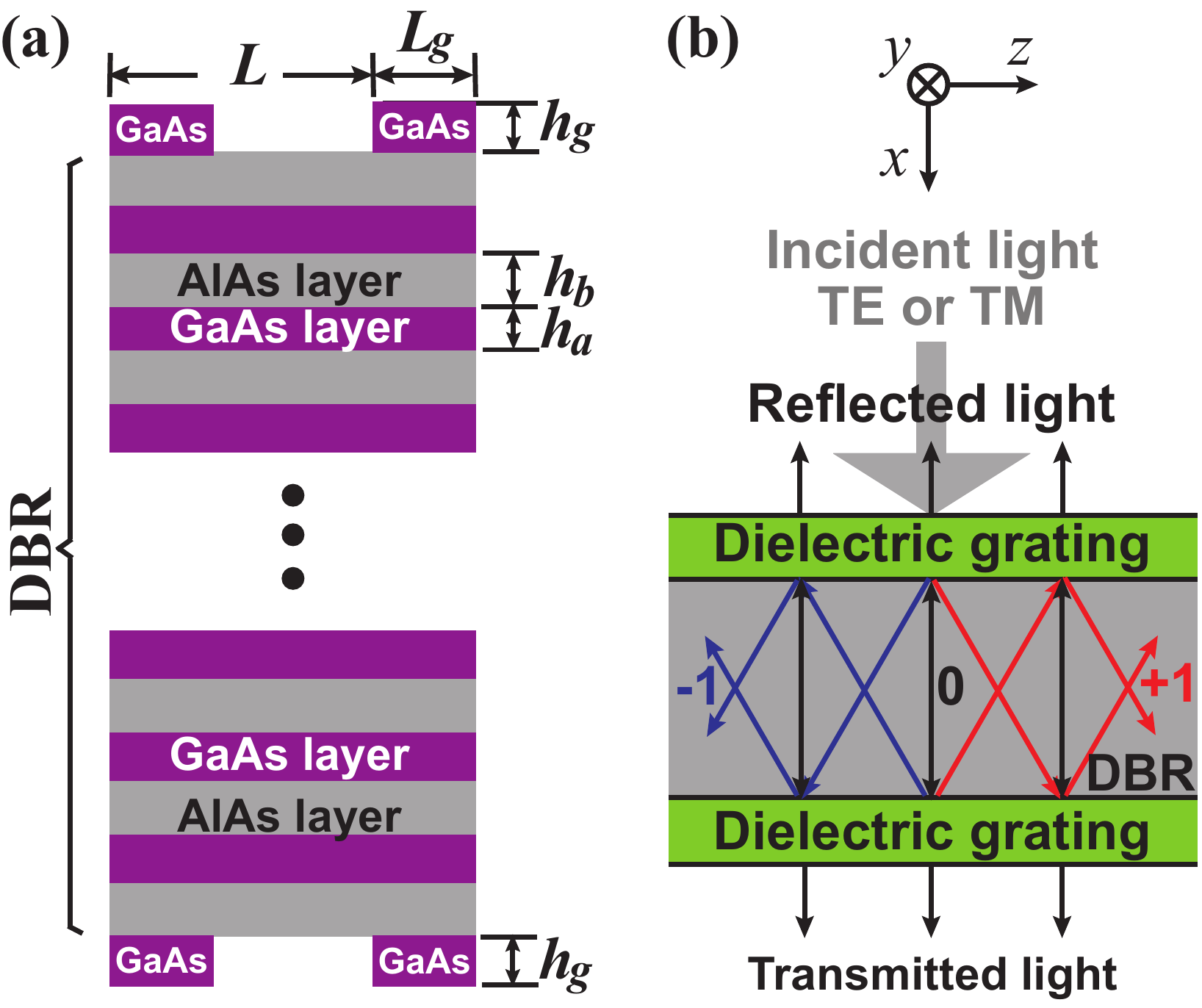}}
 \caption{\small (Color online) (a) Schematic illustration of the distribute Bragg reflector with two same GaAs gratings on top and on the bottom, respectively. (b) Geometry of diffraction waves that can transfer in (a) and the coordinate system we adopt.}
\end{figure}

The period of both GaAs gratings is set to be about 300 nm to match the DBR band gap, which implies that their reciprocal lattice vector defined as $G=2\pi/L$ is about 20 $\mu{\rm m}^{-1}$. As a result, there are only three modes, namely, the zeroth and $\pm1$ order diffraction waves, that can transfer in the DBR, referred to Fig.~1(b). Therefore, the electromagnetic wave can be expanded as their sum, namely,
\begin{align}
Y^{(j)}_y (x,z)=\sum_\mu Y^{(j)}_\mu (x)e^{i\mu Gz}
\end{align}
where $Y$ represents the $y$ component of the electric field for the TE mode or magnetic field for the TM mode. $\mu $ is 0 and $\pm1$, and $j$ is the sequence number of the layer. The coordinate system is shown in Fig.~1(b). For the $j$th layer, the coupled equations of  $ Y^{(j)}_\mu $ can be cast into
\begin{align}
\frac{\partial^2 }{\partial x^2}{\bm y}^{(j)}+{\bm \Xi}^{(j)}\cdot{\bm y}^{(j)}=0
\end{align}
where ${\bm y}^{(j)}=\left[Y^{(j)}_{-1},Y^{(j)}_{0},Y^{(j)}_{+1}\right]^T$. The matrix ${\bm \Xi}^{(j)}$ is
\begin{align}
{\bm \Xi}^{(j)}_{\rm TE}&=\left[
\begin{array}{ccc}
\varepsilon_0^{(j)}k_0^2-G^2 & \varepsilon_{-1}^{(j)}k_0^2 & \varepsilon_{-2}^{(j)}k_0^2\\
\varepsilon_1^{(j)}k_0^2 & \varepsilon_{0}^{(j)}k_0^2 & \varepsilon_{-1}^{(j)}k_0^2\\
\varepsilon_2^{(j)}k_0^2 & \varepsilon_{1}^{(j)}k_0^2 & \varepsilon_{0}^{(j)}k_0^2-G^2 \\
\end{array}
\right]
\end{align}
for the TE mode, and
\begin{align}
{\bm \Xi}_{\rm TM}^{(j)}&\!=
\!\!\left[
\begin{array}{ccc}
\eta_0^{(j)} & \eta_{-1}^{(j)} & \eta_{-2}^{(j)}\\
\eta_1^{(j)} & \eta_0^{(j)}  & \eta_{-1}^{(j)} \\
\eta_2^{(j)} & \eta_1^{(j)}  & \eta_0^{(j)}\\
\end{array}
\right]^{{-}1} \!\left[
\begin{array}{ccc}
k_0^2{-}\eta_0^{(j)}G^2 & 0 & \eta_{-2}^{(j)}G^2\\
0  &   k_0^2 & 0 \\
\eta_{2}^{(j)}G^2  & 0 &k_0^2{-}\eta_0^{(j)}G^2\\
\end{array}
\right]
\end{align}
for the TM mode. Here, 
\begin{align}
\varepsilon_\mu^{(j)}&=(\mu \pi)^{-1} \left[\sin(\mu \pi)+\left(\varepsilon^{(j)}-1\right) \sin\left(\chi^{(j)}\mu \pi\right)\right],\nonumber\\
\eta_\mu^{(j)}&=(\mu \pi)^{-1} \left[\sin(\mu \pi)+\left({1\over\varepsilon^{(j)}}-1\right) \sin\left(\chi^{(j)}\mu \pi\right)\right]\nonumber
\end{align} 
represent the $\mu$th order Fourier component of the $j$th layer permittivity $\varepsilon^{(j)}$ and $1/\varepsilon^{(j)}$, respectively. $\chi^{(j)}$ is the duty cycle of the $j$th layer, and equals $f$ for the grating layers and 1 for the rest. With the help of $\bm \Xi^{(j)}$, one can find the transfer matrix, $\bm M^{(j)}$, of $\bm V^{(j)}=\left[Y^{(j)}_{-1}, Y^{\prime(j)}_{-1}, Y^{(j)}_{0}, Y^{\prime(j)}_{0}, Y^{(j)}_{+1}, Y^{\prime(j)}_{+1}\right]^T$ for the $j$th layer. The derivation of the transfer matrix $\bm M^{(j)}$ in detail is similar to the method developed in Ref.~\cite{Dong2012Tailoring-OL}. Using $\bm M^{(j)}$, we can obtain the total transfer matrix $\bm M$ as
\begin{align}
\bm M=\bm M^{(2N+3)}\cdots \bm M^{(2)}\bm M^{(1)}
\end{align} 
and then the transmissivity $T$ and reflectivity $R$.

\begin{figure}
\centering
\includegraphics[width=8.5 cm, bb=13 0 450 260]{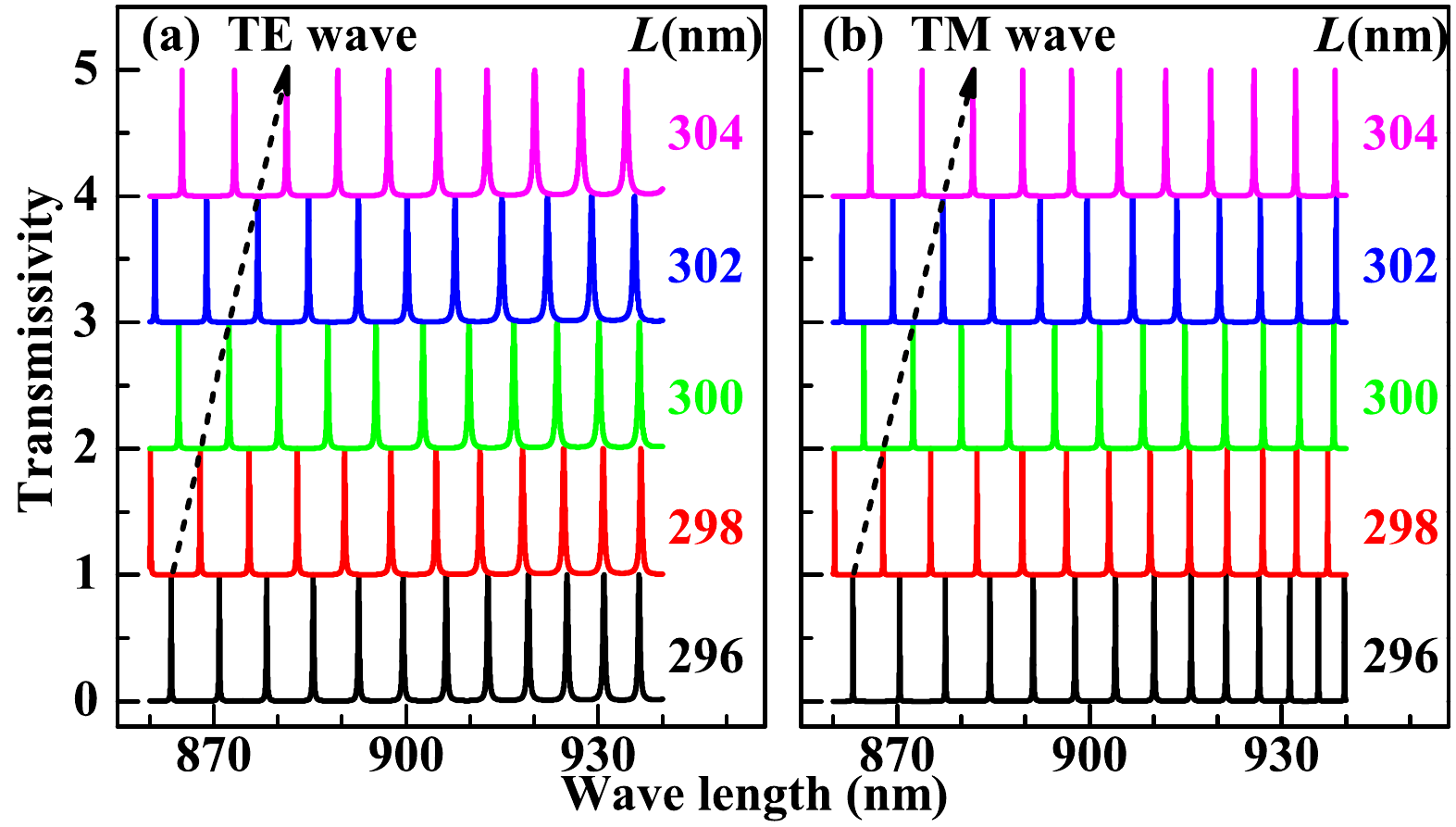}
\caption{ Transmission spectra for TE wave input (a) and TM wave input (b) under several grating periods. The dashed arrow lines in (a) and (b) guide the shift direction of the TE and TM transmission peaks, respectively, with increasing the grating period $L$. The value of $L$ is marked in two panels. Other parameters adopted are: $h_g=25$ nm, $f_g=0.5$, and {$N=48$}.}
\end{figure}

Figures 2(a) and 2(b) show the transmission spectra for the TE- and TM-wave incidence, respectively. They hold plenty of transmission peaks, and therefore appear as the transmission combs (TCs). These transmission peaks appear in the forbidden band of the DBR and are fully from the two surface gratings, referred to Fig.~1(a). From Fig.~1(b), we can see that the top grating diffracts the incident light into three kinds of modes, namely, zeroth and $\pm1$ order modes, and therefore, there are three kinds of ways for the incident light to cross the DBR. This leads to the interference of the corresponding three kinds of the optical modes and subsequently forming the TCs. The TCs for both TE and TM waves shift toward the long wavelength side with increasing the grating period $L$, displayed by the dashed arrow lines in Fig.~2. When $L$ increases, the horizontal wave vectors of the $\pm1$ order modes, being $G=2\pi/L$, decrease. Thus, the vertical wave vector of three diffraction modes should also decrease to keep their resonant interference. These lead to the redshift of the TCs when $L$ increases.

From the above consideration, we can estimate that the redshift values of the TC, $\Delta$, is about ${(n_b^2L/\lambda)\Delta L}=3.0\Delta L$ with $\Delta L$ being the $L$ increment. This implies that the relations of the TC shifts with the grating period are almost linear for the TE and TM waves, and could be fitted by
\begin{align}
\Delta_p=\eta_p\Delta L
\end{align}
where $\Delta_p$ ($p=$TE or TM) denotes the increment of the TC position as the grating period $L$ increases by $\Delta L$. The fitted parameter $\eta_p$ equals to 2.25 and 2.34 for the TE and TM waves, respectively, according to Figs.~2(a) and 2(b). The deviation of $\eta_p$ from 3.0 is mostly due to that the two gratings also create extra phase shifts for the three diffraction modes, and they are different for different frequency waves. The equation (6) indicates that the work region of the TCs could be designed by exactly controlling the grating period. In order to describe the comb-like spectra, we introduce two parameters, namely, average full width at half maximum (FWHM), ${\delta \lambda}$, of the comb peaks, and average free spectral range (FSR), $\Delta \lambda$, defined as the wavelength separation between adjacent transmission peaks. The transmission peaks we consider are limited to the wavelength range of $(860,940)$ nm.  In Fig.~2, they are about $(\delta \lambda,\Delta \lambda)=(0.30,7.19)$ nm and $(0.12,6.66)$ nm for TE and TM waves, respectively. The total height of the structure calculated in Fig.~2 is about 6.7 $\mu$m. When one use a GaAs-based semiconductor microcavity to achieve the present comb-like transmission, the height of the semiconductor microcavity should be about 20.8 $\mu$m. When their transmission lines are able to compare with each other, the height of the GaAs-based semiconductor microcavity is about three times larger than the present structure's.

\begin{figure}
  \centering
  \includegraphics[width=8.5 cm]{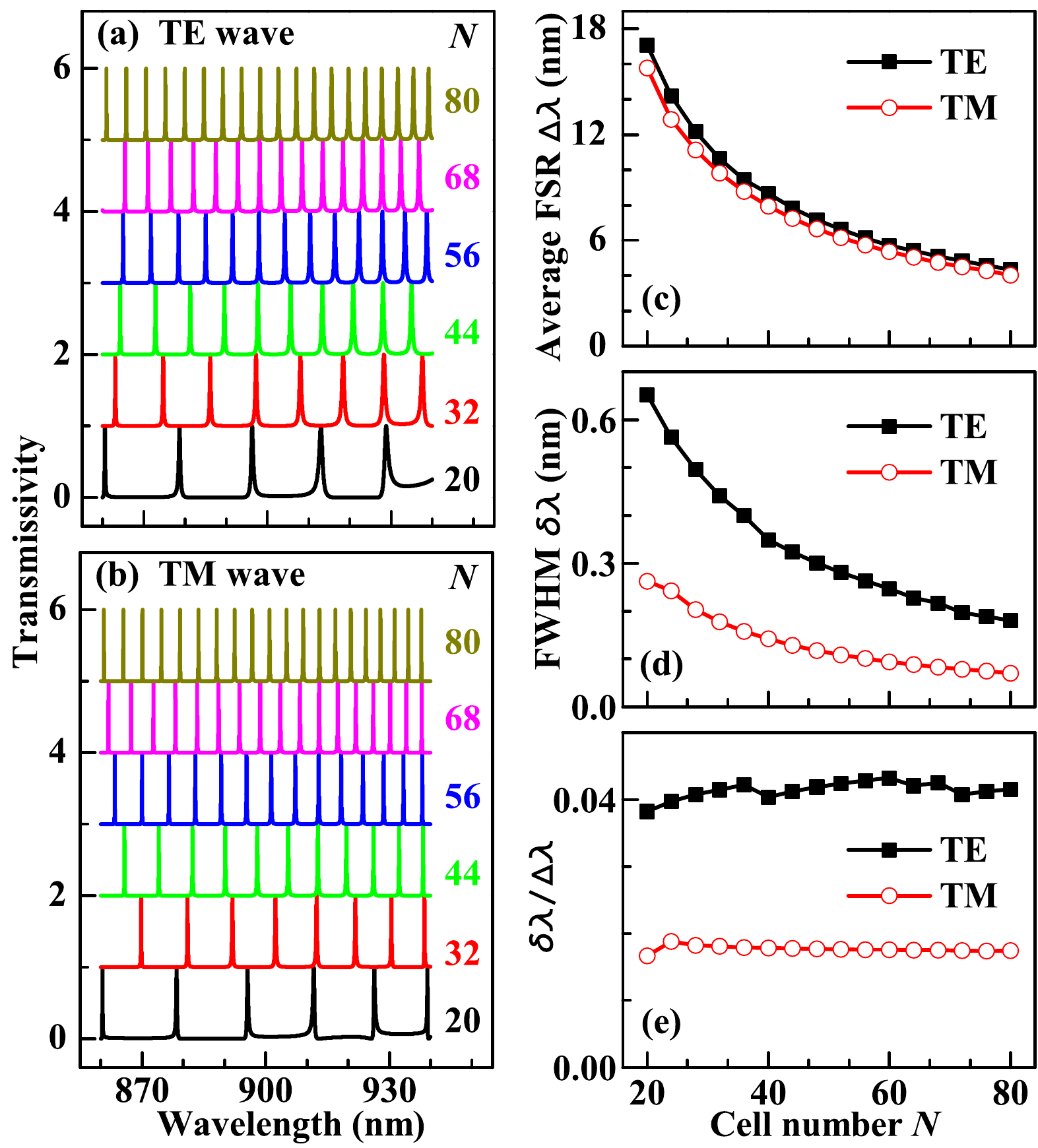}
  \caption{Transmission spectra for TE wave input (a) and TM wave input (b) under several different cell number of the DBR. In (c) and (d), the variations of the average free spectral range (FSR) and full width at half maximum (FWHM) with the cell number of the DBR are drawn, respectively, and their ratio is plotted in (e). Other parameters used are: $h_g=25$ nm, $f_g=0.5$, and {$L=300$ nm}.}
\end{figure}

Figure~3 shows the influence of the cell number $N$ on the transmission spectra of the TE and TM waves. For making this influence clear, we compare the present structure with the semiconductor microcavity again. For the semiconductor FP microcavity, $\delta \lambda$ and $\Delta \lambda$ are mainly determined by the DBR reflectivity and the cavity length, respectively. It is impossible to decrease $\delta \lambda$ and $\Delta \lambda$ simultaneously by controlling the DBR reflectivity, while this can be achieved by the present structure. From Figs.~3(a) and 3(b), we could see that the peaks become intenser and narrower when the cell number $N$ increases, corresponding to the decrease of $\Delta \lambda$ and $\delta \lambda$, respectively. The decreases of $\Delta \lambda$ and $\delta \lambda$ with increasing $N$ are shown in Figs.~3(c) and 3(d). This implies that the DBR in the present structure completes the effects of the cavity layer and two mirrors compared with the semiconductor microcavity. The average FSRs for the TE and TM transmission spectra are almost the same, which means that the TCs for the TE and TM waves hold the same peak number density, referred to Figs.~3(a) and 3(b). However, the average FWHM of the TC for the TM waves is about half of the one for the TE waves, referred to Fig.~3(d). This is due to that the TM waves hold the lower transmissivity than the TE waves during the forbidden band of the DBR.

The ratio of $\delta \lambda/\Delta \lambda$ is about 0.04 and 0.02 for the TE and TM waves, respectively, referred to Fig.~3(e). They are almost invariant with respect to the cell number $N$. The influence of the cell number $N$ on the ratio of the FWHM and FSR confirms that the DBR in the present structure plays the roles of the cavity layer and two mirrors compared with the semiconductor microcavity. The small values of $\delta\lambda$ and $\delta \lambda/\Delta\lambda$ tell us that the present structures could be taken as the candidate of the multichannel optical filter for both TE and TM waves.

\begin{figure}
  \centering
  \includegraphics[width=8.5 cm]{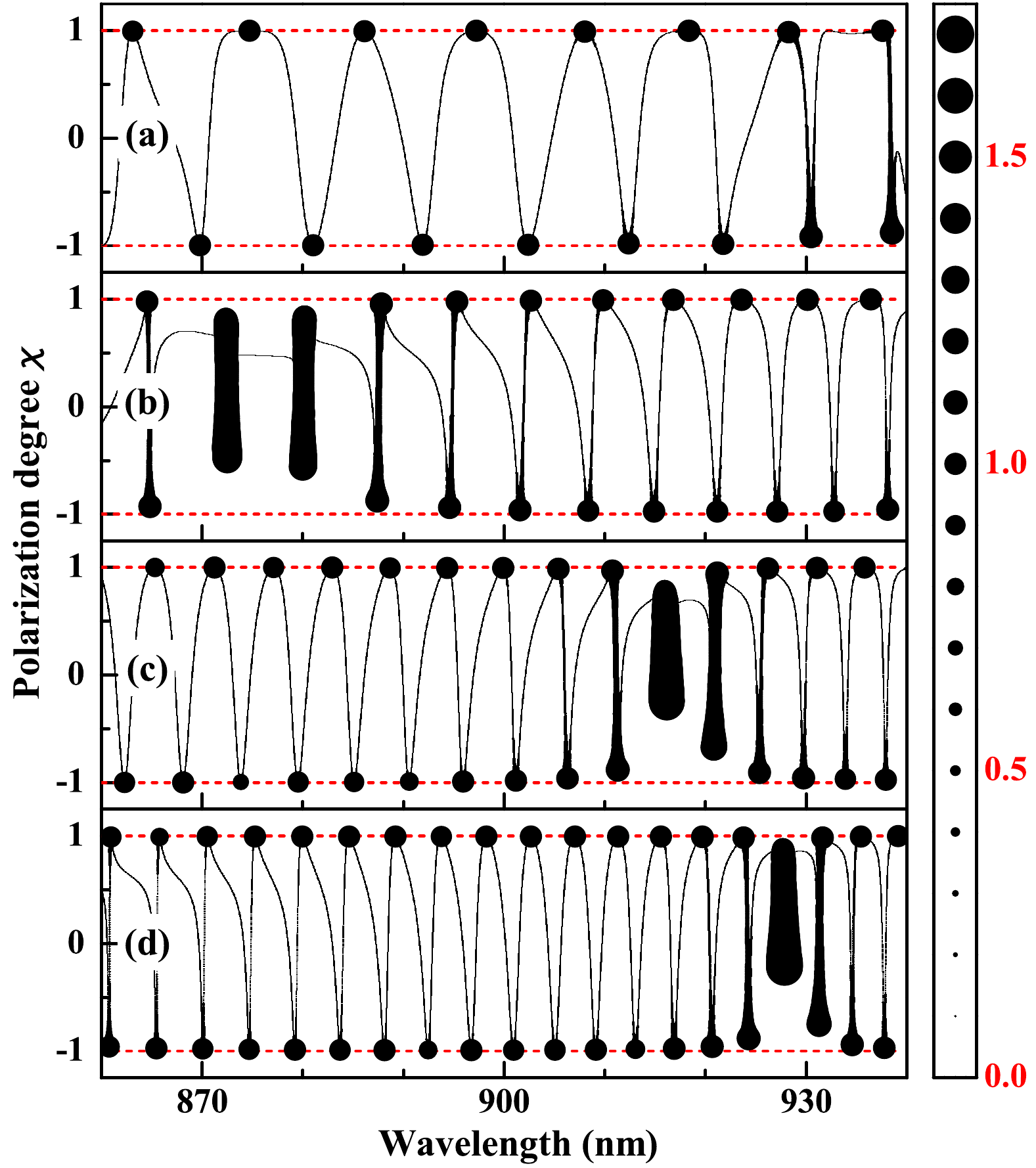}
  \caption{Variation of the transmission polarization degree with the wavelength of the incident beam. The thickness of curves represents the sum of the transmissivities of the TE and TM waves. The DBR cell number $N$ used is 32 (a), 48 (b), 64 (c), and 80 (d). Other parameters adopted in all panels are: $h_g=25$ nm, $f_g=0.5$, and {$L=300$ nm}.}
\end{figure}

Comparing the TCs of the TE and TM waves, we can find that their transmission peaks do not coincide with each other, and therefore, the present structure displays a transmission polarization. For describing it, we introduce the polarization degree, $\chi$, defined as
\begin{align}
\chi={T_{\rm TE}-T_{\rm TM}\over T_{\rm TE}+T_{\rm TM}}
\end{align}
where $T_{\rm TE}$ and $T_{\rm TM}$ are the transmissivities of the TE and TM waves, respectively. The polarization degree, shown in Fig.~4, can be up to $\pm1$ when the wavelength takes some certain values. Note that the thickness of the curves represents the sum of the transmissivities of the TE and TM waves, namely, $T_{\rm TE}+T_{\rm TM}$. The scale of their sum is given on the right side of the Fig.~4. When $\chi=\pm1$, $T_{\rm TE}$ and $T_{\rm TM}$ could reach their maximum values, displayed as the black dots in Fig.~4. The black dots for $\chi=1~(-1)$ correspond to the comb peaks of the TE (TM) transmission spectra. The black dots for $\chi=1~(-1)$ also imply that the TE (TM) comb peaks are very sharp and clearly separated from the TM (TE) wave transmission peaks, otherwise they would appear as the thick lines, referred to Fig.~4. Hence, we could expect that the present structure can serve as the multichannel polarizers during some certain wavebands. For example, the figure 4 shows that this waveband could cover almost all wavelengths, such as the bands of $(860,~930)$ nm in Fig.~4(a), of $(910,~940)$ nm in Fig.~4(b), and of $(860,~910)$ nm in Fig.~4(d).

\begin{figure}
\centering
\includegraphics[width=8.0 cm]{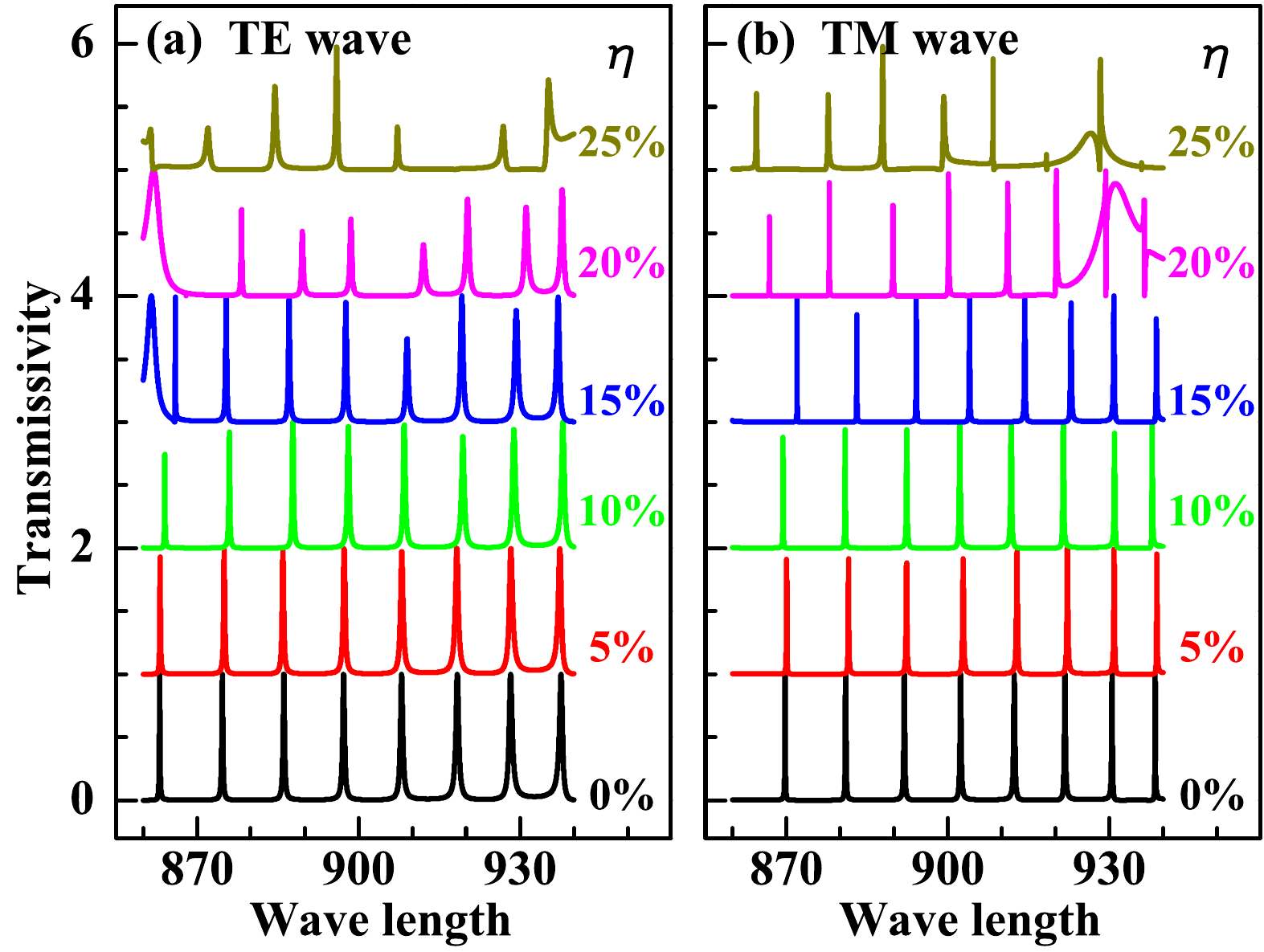} 
\caption{ Transmission spectrum of TE (a) and TM (b) waves under several randomnesses of the heights of the DBR layers. The heights of the GaAs and AlAs layers in the DBR are set to be $h_a=(1+\eta){\lambda_c/4n_a}$ and $ h_b=(1+\eta){\lambda_c/4n_b}$, respectively. Other parameters used are: $h_g=25$ nm, $f_g=0.5$, $N=32$, and {$L=300$ nm}.}
\end{figure}

From the above results, we know that the present structure holds two important applications, namely, multichannel filters and polarizers. Our further calculations also discover that these two applications show a very weak dependence on the heights and duty cycles of the two surface gratings. However, we need to note that the gratings should effectively diffract the incident beam into $\pm1$ order modes, referred to Fig.~1(b). Therefore, the grating should be high enough, simultaneously with a proper duty cycle. Our calculations indicate that the present structure can work well when $h_g>20$ nm and $f\in(0.3,0,7)$. In addition, we also calculate the dependence of the transmission comb on the randomness of the heights of the DBR layers. That is, we set 
$$h_a=(1+\eta){\lambda_c\over4n_a},~~~~h_b=(1+\eta){\lambda_c\over4n_b}$$ 
for the GaAs and AlAs layers, respectively, with $\eta$ being a normal random number. The results are shown in Fig.~5 from which we can see that the transmission spectra for the TE and TM waves still exhibit the comb-like form even when the randomness $\eta$ is up to $15\%$. Therefore, the transmission combs for the TE and TM waves are weakly dependent on the DBR defects, which displays the robustness of the transmission combs against the randomness of the heights of the DBR layers. This indicates that the present structure is in practice for experimentalists. 

To summarize, by using two surface gratings, we generated a comb-like transmission during the forbidden band of the DBR for the TE and TM incident waves. This comb-like transmission was studied by the transfer matrix approach. The size of the present structure is about one third of the one of the semiconductor microcavity when they support a similar comb-like transmission. The transmission of the present structure is polarized, and the polarization degree can be up to $\pm1$ at the TE and TM transmission peaks. Beside, the present structure is robust against the randomness of the heights of the DBR layers. Even when a 15\% randomness is introduced for them, the transmission spectrum still is comb-like. These characters guarantee a large application in optics, such as multichannel filters and polarizers. 

\bibliographystyle{aipnum4-1}
%

\end{document}